\documentclass[conference]{IEEEtran}
\IEEEoverridecommandlockouts
\usepackage{cite}
\usepackage{amsmath,amssymb,amsfonts}
\usepackage{graphicx}
\usepackage{textcomp}
\usepackage{xcolor}
\usepackage{url}
\def\BibTeX{{\rm B\kern-.05em{\sc i\kern-.025em b}\kern-.08em
    T\kern-.1667em\lower.7ex\hbox{E}\kern-.125emX}}

\newcommand{\proc}[1]{\ifmmode\mbox{\textsc{#1}}\else\textsc{#1}\fi}

\newcommand{\hema}{\proc{HEMA}}

\usepackage{algorithm}
\usepackage[noend]{algpseudocode}
\definecolor{redcolor}{rgb}{1, 0, 0}

\begin{document}

\title{HEMA: A Hands-on Exploration Platform for MEMS Sensor Attacks}

\author{\IEEEauthorblockN{Bhagawat Baanav Yedla Ravi}
\IEEEauthorblockA{\textit{Department of ECE} \\
\textit{University of Florida}\\
Gainesville, FL 32611, USA \\
b.yedlaravi@ufl.edu}
\and
\IEEEauthorblockN{Md Rafiul Kabir}
\IEEEauthorblockA{\textit{School of Engineering and Technology} \\
\textit{Central Michigan University}\\
Mount Pleasant, MI 48858, USA \\
kabir2m@cmich.edu}
\and
\IEEEauthorblockN{Sandip Ray}
\IEEEauthorblockA{\textit{Department of ECE} \\
\textit{University of Florida}\\
Gainesville, FL 32611, USA \\
sandip@ece.ufl.edu}
}

\date{}
\maketitle

\begin{abstract}
Automotive safety and security are paramount in the rapidly advancing landscape of vehicular technology. Building safe and secure vehicles demands a profound understanding of automotive systems, particularly in safety and security. Traditional learning approaches, such as reading materials or observing demonstrations, often fail to provide the practical, hands-on experience essential for developing this expertise. For novice users, gaining access to automotive-grade systems and mastering their associated hardware and software can be challenging and overwhelming. In this paper, we present a novel, affordable, and flexible exploration platform, \hema, that enables users to gain practical, hands-on insights into the security compromises of micro-electromechanical systems (MEMS) sensors, a critical component in modern ADAS systems. Furthermore, we discuss the unique challenges and design considerations involved in creating such a platform, emphasizing its role in enhancing the understanding of automotive safety and security. This framework serves as an invaluable resource for educators, researchers, and practitioners striving to build expertise in the field.


\end{abstract}


\section{Introduction}

The landscape of automotive systems has undergone a rapid and transformative evolution, primarily driven by advancements in autonomous vehicles and Advanced Driver Assistance Systems (ADAS). These cutting-edge technologies aim to improve safety, enhance driving efficiency, and redefine the overall user experience. At the heart of these innovations are sensors and actuators, which enable real-time data collection, processing, and execution of critical vehicle functions. Components like cameras, radar, LiDAR, and Inertial Measurement Units (IMUs) work in tandem with actuators to provide vehicles with the capability to sense their environment and make informed decisions. However, this increasing reliance on digital systems has also introduced a significant vulnerability: the heightened risk of cyber-attacks. Modern vehicles, now functioning as interconnected systems with embedded communication and control networks, are susceptible to malicious exploitation. For instance, a compromised IMU sensor, which is critical for navigation and stability, can severely disrupt a vehicle’s functionality, potentially leading to unsafe driving conditions. This underscores the alarming susceptibility of automotive systems to cyber threats, where even a single compromised component can jeopardize the integrity of the entire system.

Prior research \cite{trippel2017walnut, gao2023exploring, hong2022esp, zhang2023adc} has shed light on various cyber-attack vectors targeting automotive components, particularly IMU MEMS sensors, which are central to the operation of autonomous and semi-autonomous vehicles. These investigations reveal the intricate methods used to disrupt the functionality of these sensors, from electromagnetic interference (EMI) to deliberate spoofing and signal injection attacks. Despite these significant contributions, much of the existing research has been conducted by cybersecurity experts and requires a deep understanding of advanced concepts, tools, and methodologies. Consequently, the broader community, including engineers, developers, and stakeholders who are not cybersecurity specialists, often finds it challenging to grasp and simulate these attacks effectively. For novice users and non-specialists, the steep learning curve associated with understanding and replicating sensor-based cyber-attacks creates a barrier to furthering research and awareness in this domain. Addressing this gap is critical to ensuring that the knowledge of automotive cybersecurity vulnerabilities and defenses is accessible, not only to experts but also to a wider audience, fostering a collaborative approach to building safer automotive systems.

MEMS sensors consist of a micromechanical comb-like structure that forms a capacitor. Changes in motion cause this structure to change capacity, which is measured to deduce acceleration. Exploitations on these sensors include causing the structure to artificially vibrate from the acoustics of a speaker, which leads to false readings. Several research works have been done on MEMS sensor attacks \cite{khan2022security}. Trippel {\em et al.} \cite{trippel2017walnut} investigated analog acoustic injection attacks on capacitive MEMS accelerometers, revealing vulnerabilities that enable adversarial control of sensor outputs and proposing software defenses to mitigate these risks. Gao {\em et al.} \cite{gao2023exploring} developed the KITE attack, demonstrating stable and practical acoustic transduction threats to inertial sensors in multi-degree-of-freedom systems like drones. Hong {\em et al.} \cite{hong2022esp} demonstrated the stealthy acoustic attacks on MEMS gyroscopes in vehicle ESP systems and proposed a neural network-based defense strategy validated through Carsim-Simulink co-simulation.

Similarly, there have been several exploration and educational platforms for automotive security. AutoHAL, designed for ultrasonic sensors \cite{ravi2022autohal} by Ravi {\em et al.} and \cite{10.1145/3565287.3617986} by Carter {\em et al.}; HAWA \cite{10929833}, which focuses on wheel speed sensor attacks; VeCAEP, developed for V2X systems by Prakash {\em et al.} \cite{prakash2023vecaep}; and platforms for computer vision modules \cite{10043491} by Chamarthi {\em et al.}. These platforms have demonstrated the versatility and scalability of the HELP architecture, addressing a wide range of cybersecurity and safety challenges. Kabir {\em et al.} \cite{kabir2024vise} presented ViSE, a digital twin platform that facilitates the exploration of automotive functional safety and cybersecurity. However, there is no dedicated exploration and educational platform targeting MEMS sensors.

To bridge this gap, we propose an exploration platform \hema~(\underline{\textbf{H}}ands-on \underline{\textbf{E}}xploration Platform for \underline{\textbf{M}}EMS Sensor \underline{\textbf{A}}ttacks) focused on MEMS sensor attacks. This platform is designed to equip users with limited expertise in automotive cybersecurity with the practical knowledge and hands-on experience needed to understand and address these challenges. The primary goal of this platform is to provide an interactive environment for users to intuitively explore vulnerabilities, enhancing their understanding of cybersecurity threats in modern vehicles. As part of the HELP series, it aims to advance research, foster innovation, and support education in automotive safety and security. The \hema~setup uniquely offers a hands-on approach to teaching cybersecurity compromises in MEMS sensors.

\section{\hema~Approach} \label{sec:appr}
MEMS-based wheel speed sensors provide real-time data to the vehicle’s ECU, enabling precise monitoring of each wheel’s rotation. In ABS, this data helps prevent wheel lock-up by modulating brake pressure for better traction and control. In the stability control systems, the sensors work with gyroscopes and accelerometers to detect skidding or instability, allowing selective braking and power adjustments to maintain vehicle control. \hema~ allows users to observe behavior based on various parameters the user inputs. Fig. \ref{fig:hawa-arch} shows the overview of \hema's architecture. It includes the following components:

\begin{figure}
\centering
    \includegraphics[width=1\columnwidth]{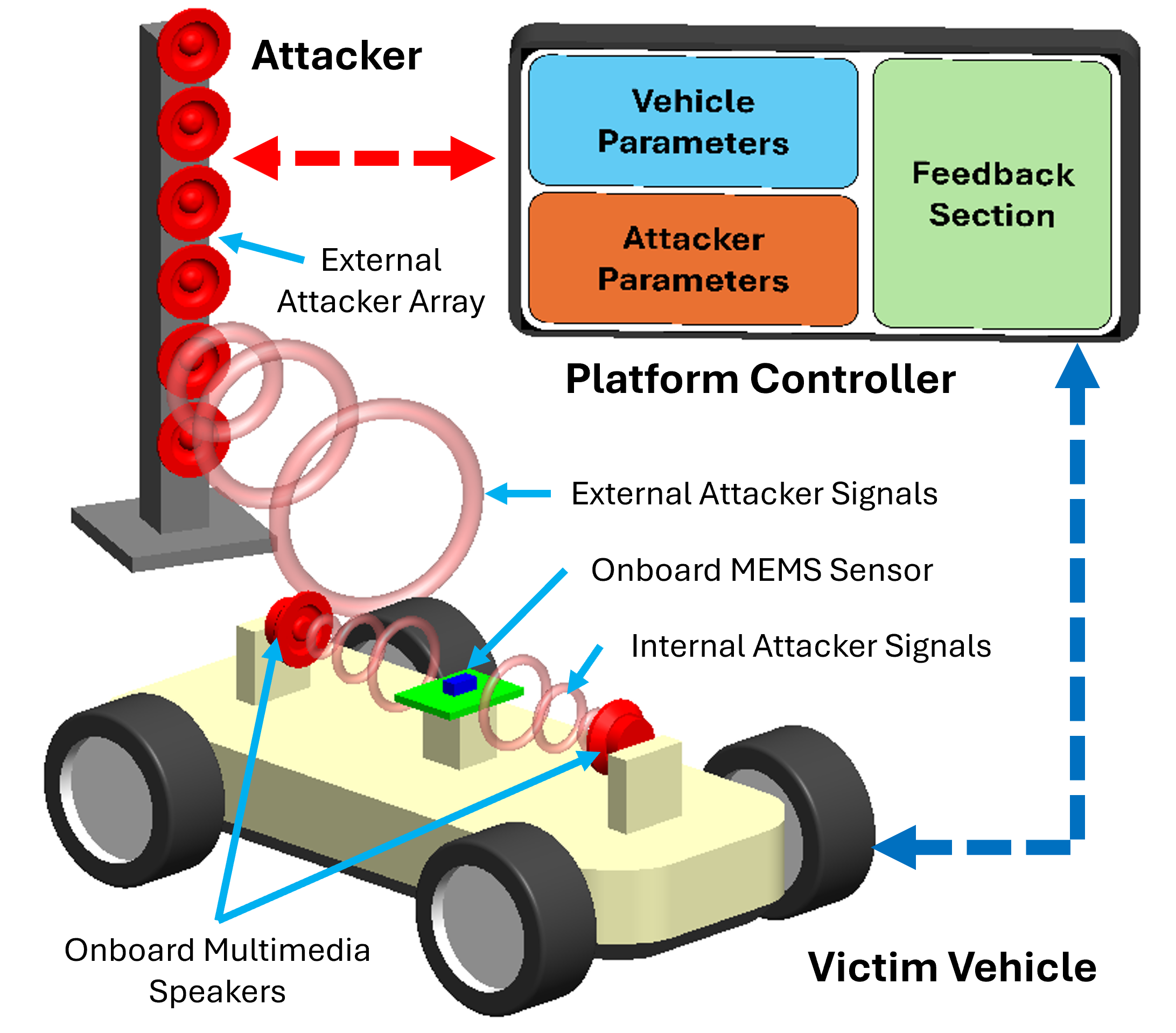}
  \caption {$\hema$ system architecture showing three entities: the victim, the attacker, and the platform controller}
  	\label{fig:hawa-arch}
\end{figure}

\subsection{Victim Vehicle}
The victim vehicle serves as the target for the attacks, allowing researchers to observe and analyze its behavior under compromised conditions. For example, the vehicle in a benign scenario would follow a trajectory defined by its parameters. The vehicle consists of a microcontroller that manages the vehicle's longitudinal and lateral motions. A MEMS sensor located in the middle of the chassis measures movement and maintains the desired speed and heading. Any discrepancy in the functioning of the MEMS sensor leads to deviation from the desired behavior. There is also a wheel speed sensor that continuously corrects the drift error from the MEMS sensor and provides ground truth information about the kinematics.

\subsection{Attacker}
It enables users to generate malicious audio signals specifically targeting onboard MEMS sensors. Research has demonstrated that MEMS sensors can be compromised through acoustic attacks, either via internal agents, such as a multimedia speaker within the system, or external agents, such as a speaker or tweeter mounted on an external entity near the vehicle. The tool provides users with the flexibility to select the type of attacker they wish to simulate, whether internal or external. Additionally, users can customize the attacker's parameters to fine-tune the attack, allowing for precise control over the nature and intensity of the threat.

\subsection{Platform Controller }
\begin{figure}
\centering
    \includegraphics[width=.99\columnwidth]{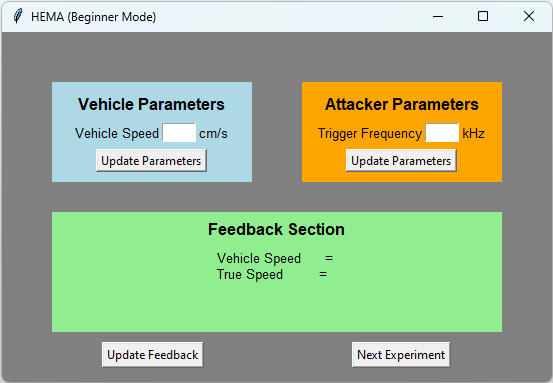}
  \caption {$\hema$ user interface}
  	\label{fig:gui}
\end{figure}

The central coordinator between the victim vehicle and the attacker is served by the platform controller. It allows users to modify vehicle parameters such as speed and heading, as well as attacker parameters like audio frequency and trigger rate. The controller includes a GUI (see Fig. \ref{fig:gui}) where users input these parameters and provide a feedback section displaying quantified data about the vehicle's behavior, enabling evaluation of the attack's outcome and effectiveness. Additionally, the controller captures malicious data from the MEMS sensor and compares it with ground truth data from the wheel speed sensor, offering initial insights into the attack's impact. It also records the dynamic behavior of the vehicle, helping users understand the overall outcome. Advanced users have the option to delve deeper by modifying the software program, enabling them to craft and explore more complex attacks creatively.

The HEMA platform is designed to meet these requirements by providing users with effective feedback during failed attempts while encouraging continued exploration of the attack. It allows users to experiment with a wide range of parameter values, including those that may not lead to a successful security compromise or produce well-defined behavior in the target vehicle. Moreover, the platform offers extensive configurability and extensibility, making it adaptable for a variety of attacks and reusable for diverse cybersecurity scenarios. This flexibility ensures that the HEMA platform fosters an interactive and iterative learning environment, enabling users to gain deeper insights into automotive security challenges.

\section{Use Case Scenario} \label{sec:usecase}
\begin{figure}
\centering
    \includegraphics[width=.9\columnwidth]{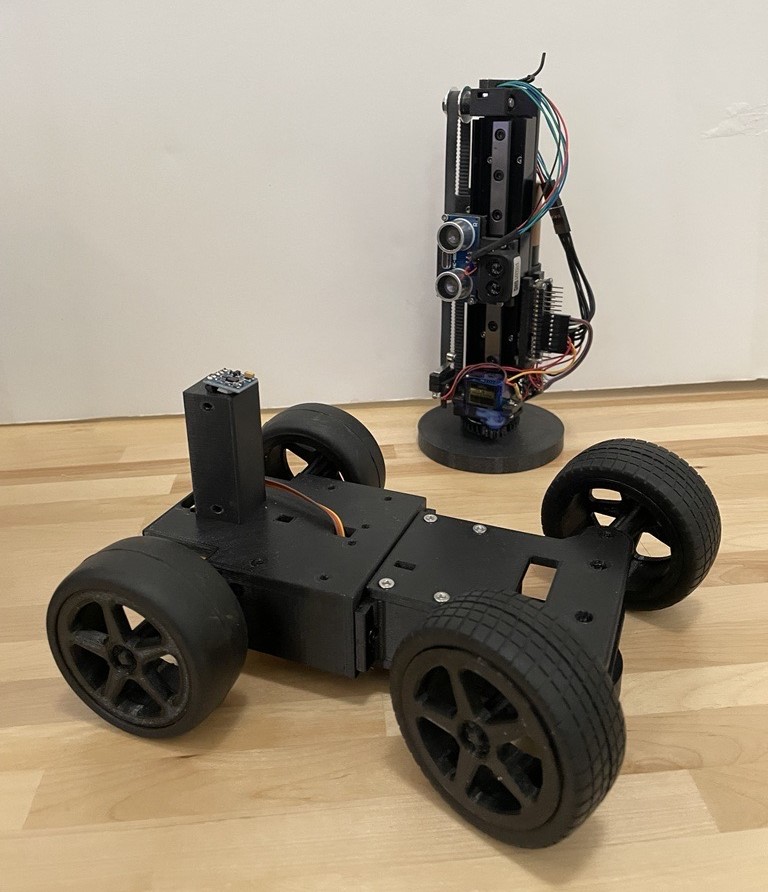}
  \caption {Physical implementation}
  	\label{fig:p}
\end{figure}

The HEMA platform demonstrates its utility through a variety of use case scenarios, focusing on both benign and attack conditions. These scenarios allow users to understand the dynamics of vehicle behavior under different operational states and evaluate the impact of malicious interference on system performance. Fig. \ref{fig:p} shows the physical implementation.
Below, we outline a representative use case to illustrate the HEMA platform's capabilities.

\subsection{Benign Scenario: Stable Vehicle Operation}
In a benign scenario, the vehicle operates as intended, following a predefined trajectory and maintaining a constant velocity as set by the user in the GUI. The user specifies the desired speed, and the vehicle accelerates until it reaches this target velocity. A PID controller ensures that the vehicle maintains this velocity by using feedback from the onboard Inertial Measurement Unit (IMU).

\subsubsection{Operational Flow}
 The user inputs the desired speed (e.g., 50 km/h) via the GUI.
The vehicle accelerates to the target speed and stabilizes using the feedback loop from the IMU.
The IMU continuously measures the vehicle's motion parameters, including velocity and heading, and sends this information to the control loop.
If there are no external disturbances, the vehicle maintains its trajectory and velocity without deviations.

\subsubsection{Outcome}
 The vehicle performs smoothly, with stable longitudinal and lateral motion. The interaction between the MEMS sensor, the wheel speed sensor, and the PID controller ensures precise kinematic control. The wheel speed sensor serves as a secondary validation mechanism, ensuring ground truth data about the vehicle’s behavior matches the IMU feedback.

\subsection{Attack Scenario: Disrupted Vehicle Operation}
In an attack scenario, a malicious actor generates acoustic interference to target the onboard MEMS sensor. The attacker’s parameters, such as audio frequency and intensity, are configured via the GUI. These attacks compromise the integrity of the IMU data, resulting in abnormal vehicle behavior.

\subsubsection{Operational Flow}
 The attacker launches a malicious acoustic signal, simulating interference from an external source (e.g., a speaker positioned near the vehicle) or an internal source (e.g., an embedded multimedia speaker).
The IMU begins to produce false data due to the acoustic attack, causing incorrect velocity feedback to the control loop.
The vehicle experiences erratic motion, such as jerks, non-uniform acceleration, or deviation from the desired trajectory.

\subsubsection{Impact}
The vehicle's motion becomes unstable, failing to maintain the user-defined velocity.
Erroneous IMU data leads to inappropriate control decisions, affecting the overall driving performance.
The wheel speed sensor detects discrepancies between its readings and the IMU data, but without effective countermeasures, the attack impacts the vehicle’s operational safety.

\subsection{Future Enhancements}
The HEMA platform will incorporate sensor fusion to detect tire slip by comparing data from the IMU and wheel speed sensor.

\subsubsection{Benign Condition}
 Accurate detection triggers the ABS to prevent wheel lock-up, maintain traction, and ensure optimal stopping distance.
\subsubsection{Attack Condition}
 Acoustic interference on the IMU leads to false readings, degrading ABS performance and increasing stopping distance.
To address limitations in simulating real-world tire slip, future upgrades will include a larger-scale vehicle with sufficient inertia, enabling realistic tests of ABS performance and skidding behavior.

\section{Educational and Training Framework}
The HEMA platform is designed not only as a tool for exploring MEMS sensor attacks but also as a valuable resource for education and professional training. Its versatility and hands-on approach make it an ideal candidate for integration into academic curricula, industry workshops, and certification programs focused on automotive security.

\subsection{Academic Integration}

HEMA can be integrated into undergraduate and graduate courses on automotive systems, embedded systems, and cybersecurity. It offers a hands-on complement to theory, allowing students to simulate real-world scenarios and study MEMS sensor vulnerabilities in critical automotive systems.

The platform can also support capstone or final-year projects, enabling students to explore attack strategies, mitigation techniques, or extend HEMA’s functionality to other subsystems like radar or LiDAR.

\subsection{Industry Training}

 HEMA can be used to conduct targeted workshops aimed at upskilling industry professionals who may have expertise in automotive system design but lack experience in cybersecurity. By simulating MEMS sensor attacks, the platform bridges the gap between system functionality and security considerations.
 
 Professionals can use HEMA to explore and mitigate attack scenarios relevant to their specific domain, such as ADAS or autonomous vehicles, thereby enhancing their understanding of how to secure modern automotive systems.

\subsection{Collaboration with Institutions}
Educational institutions and industry leaders can collaborate to deploy HEMA in training programs, tailoring it to address region-specific challenges or emerging trends in automotive security. HEMA can serve as a platform for collaboration among researchers, students, and professionals, encouraging the exchange of ideas and best practices in automotive safety and security.

\section{Conclusion} \label{sec:cncl}

MEMS sensors in the automotive field play a pivotal role in providing critical data for safety, stability, and performance, serving as the backbone for systems like ABS, ESP, and ADAS. These sensors enable real-time monitoring and control by measuring parameters such as acceleration, angular velocity, and vibration, which are essential for maintaining vehicle functionality and ensuring passenger safety. However, their integration into increasingly connected automotive systems also exposes them to potential security vulnerabilities. 
In this paper, we present a platform, \hema, designed to enable users to explore and understand attacks on MEMS sensors. \hema~ serves as an educational and training tool, offering a hands-on approach to automotive security. It is particularly valuable in two scenarios: 1) teaching students who are new to automotive systems and security concepts, and 2) training industry professionals who are familiar with automotive system functionality but may lack expertise in cybersecurity. By allowing users to simulate various attack scenarios, \hema~ provides an interactive learning environment that bridges the gap between theoretical knowledge and practical application.

In future work, we plan to extend \hema~ to facilitate the exploration and analysis of more sophisticated attack vectors, such as signal injection, acoustic interference, and electromagnetic attacks. Future iterations will incorporate simulations of the dynamic consequences of these attacks, providing insights into how they impact real-world vehicle behavior, such as stability control, braking, and navigation systems.

\bibliographystyle{IEEEtran}
\bibliography{IEEEexample}

\begin{thebibliography}{10}
\providecommand{\url}[1]{#1}
\csname url@samestyle\endcsname
\providecommand{\newblock}{\relax}
\providecommand{\bibinfo}[2]{#2}
\providecommand{\BIBentrySTDinterwordspacing}{\spaceskip=0pt\relax}
\providecommand{\BIBentryALTinterwordstretchfactor}{4}
\providecommand{\BIBentryALTinterwordspacing}{\spaceskip=\fontdimen2\font plus
\BIBentryALTinterwordstretchfactor\fontdimen3\font minus \fontdimen4\font\relax}
\providecommand{\BIBforeignlanguage}[2]{{%
\expandafter\ifx\csname l@#1\endcsname\relax
\typeout{** WARNING: IEEEtran.bst: No hyphenation pattern has been}%
\typeout{** loaded for the language `#1'. Using the pattern for}%
\typeout{** the default language instead.}%
\else
\language=\csname l@#1\endcsname
\fi
#2}}
\providecommand{\BIBdecl}{\relax}
\BIBdecl

\bibitem{trippel2017walnut}
T.~Trippel, O.~Weisse, W.~Xu, P.~Honeyman, and K.~Fu, ``Walnut: Waging doubt on the integrity of mems accelerometers with acoustic injection attacks,'' in \emph{2017 IEEE European symposium on security and privacy (EuroS\&P)}.\hskip 1em plus 0.5em minus 0.4em\relax IEEE, 2017, pp. 3--18.

\bibitem{gao2023exploring}
M.~Gao, L.~Zhang, L.~Shen, X.~Zou, J.~Han, F.~Lin, and K.~Ren, ``Exploring practical acoustic transduction attacks on inertial sensors in mdof systems,'' \emph{IEEE Transactions on Mobile Computing}, vol.~23, no.~5, pp. 3539--3557, 2023.

\bibitem{hong2022esp}
Z.~Hong, X.~Li, Z.~Wen, L.~Zhou, H.~Chen, and J.~Su, ``Esp spoofing: Covert acoustic attack on mems gyroscopes in vehicles,'' \emph{IEEE Transactions on Information Forensics and Security}, vol.~17, pp. 3734--3747, 2022.

\bibitem{zhang2023adc}
J.~Zhang, Y.~Wang, Y.~Tu, S.~Rampazzi, Z.~Lin, I.~Lee, and X.~Hei, ``Adc-bank: Detecting acoustic out-of-band signal injection on inertial sensors,'' in \emph{International Conference on Security and Privacy in Cyber-Physical Systems and Smart Vehicles}.\hskip 1em plus 0.5em minus 0.4em\relax Springer, 2023, pp. 53--72.

\bibitem{khan2022security}
A.~A. Khan, K.~Sahebkar, C.~Xi, M.~M. Tehranipoor, R.~F. Need, and N.~Asadizanjani, ``Security challenges of mems devices in hi packaging,'' in \emph{2022 IEEE 72nd Electronic Components and Technology Conference (ECTC)}.\hskip 1em plus 0.5em minus 0.4em\relax IEEE, 2022, pp. 2321--2327.

\bibitem{ravi2022autohal}
B.~B.~Y. Ravi, M.~R. Kabir, N.~Mishra, S.~Boddupalli, and S.~Ray, ``Autohal: An exploration platform for ranging sensor attacks on automotive systems,'' in \emph{2022 IEEE International Conference on Consumer Electronics (ICCE)}.\hskip 1em plus 0.5em minus 0.4em\relax IEEE, 2022, pp. 1--2.

\bibitem{10.1145/3565287.3617986}
J.~Carter, B.~B. Yedla~Ravi, M.~R. Kabir, and S.~Ray, ``Poster: Efficient exploration of automotive ranging sensor attacks,'' in \emph{Proceedings of the Twenty-Fourth International Symposium on Theory, Algorithmic Foundations, and Protocol Design for Mobile Networks and Mobile Computing}, ser. MobiHoc '23.\hskip 1em plus 0.5em minus 0.4em\relax Association for Computing Machinery, 2023.

\bibitem{10929833}
B.~B.~Y. Ravi and S.~Ray, ``Hands-on exploration and learning platform for automotive wheel speed sensor attacks,'' in \emph{2025 IEEE International Conference on Consumer Electronics (ICCE)}, 2025, pp. 1--6.

\bibitem{prakash2023vecaep}
D.~M. Prakash, B.~B.~Y. Ravi, S.~Boddupalli, and S.~Ray, ``Vecaep: A hands-on exploration platform for vehicular communication attacks,'' in \emph{2023 IEEE 97th Vehicular Technology Conference (VTC2023-Spring)}.\hskip 1em plus 0.5em minus 0.4em\relax IEEE, 2023, pp. 1--5.

\bibitem{10043491}
V.~S. Gireesh~Chamarthi, X.~Chen, B.~B. Yedla~Ravi, and S.~Ray, ``Exploration of machine learning attacks in automotive systems using physical and mixed reality platforms,'' in \emph{2023 IEEE International Conference on Consumer Electronics (ICCE)}, 2023, pp. 1--4.

\bibitem{kabir2024vise}
M.~R. Kabir and S.~Ray, ``Vise: Digital twin exploration for automotive functional safety and cybersecurity,'' \emph{Journal of Hardware and Systems Security}, pp. 1--12, 2024.

\end{thebibliography}

\end{document}